\documentclass[conference]{IEEEtran}
\IEEEoverridecommandlockouts

\usepackage{amsmath,amssymb,amsfonts}
\usepackage{graphicx}
\usepackage{textcomp}
\usepackage{xcolor}
\usepackage{booktabs}
\usepackage{multirow}
\usepackage{cite}
\usepackage{microtype}
\usepackage{array}

\usepackage{amssymb}
\begin{document}

\title{Lightweight PCGAE-Net: Parallel CrossGate Attention and Bottleneck AutoEncoder for Efficient 5G Channel Prediction}

\author{%
		\IEEEauthorblockN {Uma Kishore Godavarti\textsuperscript{1, 2}, K. Giridhar\textsuperscript{2}, Vanani Prince Dharmendrabhai\textsuperscript{1}, \\ Anchit Panday\textsuperscript{1}, and Madhan Raj Kanagarathinam\textsuperscript{1}}

		\IEEEauthorblockA{ 
			\textsuperscript{1} Network Modem Team, Samsung R\&D India Bangalore\\
                \textsuperscript{2} Department of Electrical Engineering, Indian Institute of Technology Madras, Chennai, India
		}	
	\vspace{-5ex}
    }

\maketitle

\begin{abstract}
Accurate channel state information (CSI) prediction is essential for proactive beamforming and resource management in 5G massive MIMO systems, yet the deployment of high-accuracy transformer-based predictors on base-station hardware remains challenging because the most capable models carry upwards of 30\,M parameters. This paper introduces Lightweight PCGAE-Net, which addresses the efficiency problem not by post-hoc compression but by correcting two architectural flaws in the current state of the art. The first is a sequential attention ordering bias: in CS3T-UNet, group-wise temporal attention (GTA) always operates on features that have already been transformed by cross-shaped spatial attention (CSA), distorting what temporal information GTA can capture. We remove this dependency by routing both attention modules to the same layer-normalized input and combining their independent outputs through a learned per-channel sigmoid CrossGate. The second flaw is an uncompressed bottleneck: applying full self-attention at the deepest encoder stage, where channel depth reaches $4C$, is quadratically expensive and carries redundant features. A Bottleneck AutoEncoder (BAE) with $1\times1$ convolutions halves this depth and uses an auxiliary reconstruction loss to prevent information collapse. Wrapping these components inside a shallower encoder-decoder with frequency-domain dimensionality reduction ($N_f\!=\!32$, $C\!=\!48$) produces a model with just 8.54\,M parameters---58\% fewer than the CS3T-UNet baseline---that outperforms it by up to 3.26\,dB at 5\,km/h and 6.0\,dB at 9\,km/h in single-step prediction on QuaDriGa dataset.
\end{abstract}


\section{Introduction}

The ability to predict future channel conditions from past observations has become an important signal processing step in 5G and beyond 5G systems. In massive MIMO systems, the base station needs accurate  CSI to compute beamforming weights, but obtaining that information requires pilot transmission, feedback, and processing time that consume the coherence window~\cite{truong2013channel}. A predictor that can accurately forecast the channel one or several time steps ahead, using only the history of observed frames, lets the base station prepare precoding matrices in advance and tolerate feedback delays that would otherwise degrade throughput~\cite{love2008overview}.

For several years, recurrent architectures were the natural choice for this task. LSTM and GRU networks can model sequential temporal evolution, and their predictions improve noticeably over simple extrapolation~\cite{ye2018dl}. However, recurrent models process time steps one at a time, which limits parallelism and makes it hard to capture long-range dependencies across a channel sequence of tens of frames. CNN-based approaches~\cite{wen2018csinet} avoid this by treating the channel matrix as a 2D spatial image, but trade temporal modeling for spatial feature extraction rather than handling both jointly.

Transformers changed this picture. The self-attention mechanism can attend to arbitrary positions in both spatial and temporal dimensions simultaneously, and several recent channel prediction systems have used transformer-based encoder-decoder architectures to reach state-of-the-art NMSE~\cite{vaswani2017attention,yuan2021transformer}. CS3T-UNet~\cite{cs3tunet2024} is among the most capable, combining a U-Net backbone with cross-shaped spatial attention (CSA) and group-wise temporal attention (GTA). At 20.34\,M parameters, it achieves strong results, but two aspects of its design limit both performance and efficiency.

The first is the order in which it applies attention. In CS3T-UNet, CSA runs first and GTA is applied to its output, meaning GTA always sees a spatially-transformed version of the channel features. When spatial and temporal dynamics are partially coupled---as they are in MIMO systems where physical angle-of-arrival varies over time---applying spatial processing before temporal processing biases what GTA can extract. The model partially compensates during training, but the structural constraint remains. The second issue is the bottleneck: at the deepest encoder layer, feature depth reaches $4C$ (Where C is Embedding dimension), and CS3T-UNet applies cross-shaped attention at full resolution. Self-attention cost scales quadratically with channel depth, so this is the most expensive operation in the network, applied to a feature volume that carries unavoidable redundancy at such depth.

These are architectural problems, not weight-level inefficiencies, so post-hoc techniques like pruning or quantization cannot resolve them. The clean solution is to redesign the attention mechanism. This paper presents Lightweight PCGAE-Net, which makes the following specific contributions:

\begin{enumerate}
\item \textbf{Parallel CrossGate Attention Block:} CSA and GTA are computed simultaneously on identical layer-normalized inputs, removing the ordering dependency. Their outputs are combined by a learned per-channel sigmoid gate that adapts the spatial-temporal mixture based on the actual feature content, with zero-initialization ensuring a balanced starting point.

\item \textbf{Bottleneck AutoEncoder with Auxiliary Supervision:} At the deepest encoder stage, $1\times1$ convolutions compress feature depth from $4C$ to $2C$, cutting attention cost by roughly 75\%. An auxiliary reconstruction loss prevents the compressed bottleneck from discarding channel structure needed by the decoder.

\item \textbf{Slim Encoder-Decoder and Delay Domain Dimensionality Reduction:} Reducing the number of merge/expand block pairs from three to two, and cropping the input to $N_f\!=\!32$ most energetic ADP delay bins with base channel width $C\!=\!48$, brings total model size to 8.54\,M parameters---73\% fewer than the full PCGAE-Net and 58\% fewer than CS3T-UNet.

\item \textbf{Systematic evaluation:} Five model variants are evaluated at three UE speeds (5, 9, 12\,km/h) and two prediction horizons ($L\!\in\!\{1,5\}$) on QuaDriGa~\cite{jaeckel2014quadriga} under 3GPP 38.901 UMa NLOS conditions, with ablation for each design choice.
\end{enumerate}

The paper is organized as follows. Section~\ref{sec:related} reviews related work. Section~\ref{sec:model} describes the system model. Section~\ref{sec:arch} presents the proposed architecture. Section~\ref{sec:exp} details the experimental setup. Section~\ref{sec:results} reports results and analysis. Section~\ref{sec:conclusion} concludes.

\section{Related Work}
\label{sec:related}

\subsection{Channel Prediction with Deep Learning}

The deep learning work on channel prediction adapted models developed for time-series forecasting. LSTM networks were applied to OFDM channel matrices by treating the sequence of channel observations as a multivariate time series~\cite{ye2018dl}. This works well at low antenna counts but does not scale well to massive MIMO, where the spatial dimension has hundreds of entries, and the spatial structure is as informative as the temporal one. GRU-based approaches~\cite{lu2021cpr} offer a lighter recurrent alternative and remain competitive at low mobility.

CNNs entered through the CSI feedback compression literature~\cite{wen2018csinet}, where autoencoders learned to compress and reconstruct channel matrices. The encoder half naturally learns spatial feature representations, and it was a short step to reuse those representations for prediction. The limitation is that CNNs have fixed receptive fields and cannot model dependencies between frames that are far apart in time without deep architectures that carry correspondingly large parameter counts.

Transformers solved this by replacing convolutions with self-attention, which has a global receptive field by construction~\cite{vaswani2017attention}. Several groups have since applied transformer architectures to channel prediction, generally following the pattern of a spatial feature extractor followed by a temporal attention module~\cite{yuan2021transformer}. CS3T-UNet~\cite{cs3tunet2024} achieves the best published NMSE on QuaDriGa dataset by combining CSA with GTA inside a U-Net encoder-decoder. Our work retains this backbone structure, but restructures the attention to run in parallel rather than in sequence, and adds a compressed bottleneck.

\subsection{Efficient Transformer Design}

The tension between transformer expressiveness and computational cost has received considerable attention in vision. Swin Transformer~\cite{liu2022swin} addressed this by partitioning the image into local windows and computing attention only within each window, which reduces complexity from quadratic in total token count to quadratic in window size. The cross-shaped attention in CS3T-UNet follows a similar logic: each token attends only to positions in the same horizontal or vertical strip. We retain this efficient spatial attention design; our contribution addresses how spatial and temporal attentions interact rather than how each handles long-range spatial dependencies.

For general-purpose efficient network design, MobileNet~\cite{howard2017mobilenets} and EfficientNet~\cite{tan2019efficientnet} showed that architectures designed for efficiency from the start outperform those compressed after training. Depthwise separable convolutions, width multipliers, and compound scaling rules exemplify this design-for-efficiency philosophy. We apply the same principle to transformer-based channel prediction, accepting a small accuracy cost in exchange for a substantially smaller model rather than training a large accurate model and then compressing it~\cite{ravindran2023lightweight}.

\subsection{CSI Compression and Representation Learning}

Autoencoder-based CSI compression~\cite{wen2018csinet,wang2019csifeedback} has been studied extensively for FDD feedback, where the encoder at the UE compresses CSI into a bitstream and the decoder at the base station reconstructs it. The reconstruction quality is the primary objective in that setting. Our Bottleneck AutoEncoder differs in two important respects. It operates inside the prediction network rather than between the transmitter and receiver, and its reconstruction loss is a regularizer rather than the primary training objective. The goal is not to faithfully reconstruct the compressed features but to prevent the compressed representation from losing channel structure needed for an accurate prediction. This distinction changes the compression ratio appropriate and how the auxiliary loss should be weighted, as we discuss in Section~\ref{sec:arch}.

\section{System Model}
\label{sec:model}

\subsection{Channel Model}

We consider a massive MIMO downlink in which a base station with $N_t\!=\!64$ antennas serves a single-antenna UE that moves at speed $v$. The uplink channel at discrete time step $t$ is $\mathbf{H}(t)\!\in\!\mathbb{C}^{N_f\times N_t}$, where $N_f\!=\!64$ is the number of OFDM subcarriers. Each entry $[\mathbf{H}(t)]_{k,m}$ is the complex channel gain between UE and base station at each antenna $m$ and subcarrier $k$.

The raw channel matrices are correlated in both frequency and spatial dimensions. To exploit this structure, we transform to the angle-delay-power (ADP) domain:
\begin{equation}
\mathbf{H}_{\mathrm{ADP}}(t) = \mathbf{F}_f\,\mathbf{H}(t)\,\mathbf{F}_t^H,
\label{eq:adp}
\end{equation}
where $\mathbf{F}_f\!\in\!\mathbb{C}^{N_f\times N_f}$ and $\mathbf{F}_t\!\in\!\mathbb{C}^{N_t\times N_t}$ are DFT matrices applied along frequency and spatial dimensions, respectively. In the ADP domain, channel energy concentrated in a small number of dominant angle-delay bins corresponding to physical propagation paths. For typical urban macro scenarios, most energy sits in 15--20\% of the bins. Truncating to the $N_f\!=\!32$ most energetic delay indices retains nearly all channel information while halving the effective input size---the basis for our delay domain dimensionality reduction strategy.

\subsection{Prediction Task Formulation}

The predictor at the base station observes $T\!=\!10$ consecutive ADP-domain frames and outputs predictions for the following $L$ steps. Stacking observations along the time axis gives an input tensor $\mathcal{X}\!\in\!\mathbb{R}^{T\times 2\times N_f\times N_t}$, where the factor of 2 separates the real and imaginary parts. The prediction target is $\mathcal{Y}\!\in\!\mathbb{R}^{L\times 2\times N_f\times N_t}$.

Prediction accuracy is reported as normalized mean-squared error:
\begin{equation}
\mathrm{NMSE} = 10\log_{10}\!\left(\frac{\|\hat{\mathcal{Y}}-\mathcal{Y}\|_F^2}{\|\mathcal{Y}\|_F^2}\right),
\label{eq:nmse}
\end{equation}
where $\hat{\mathcal{Y}}$ denotes the model output, and lower values indicate better accuracy. For practical gNB deployment, the inference time per slot must fit within the channel coherence time, which places a hard upper bound on model complexity and motivates our parameter-reduction strategy.

\section{Proposed Architecture}
\label{sec:arch}

\begin{figure*}[t]
\centering
\includegraphics[width=\textwidth]{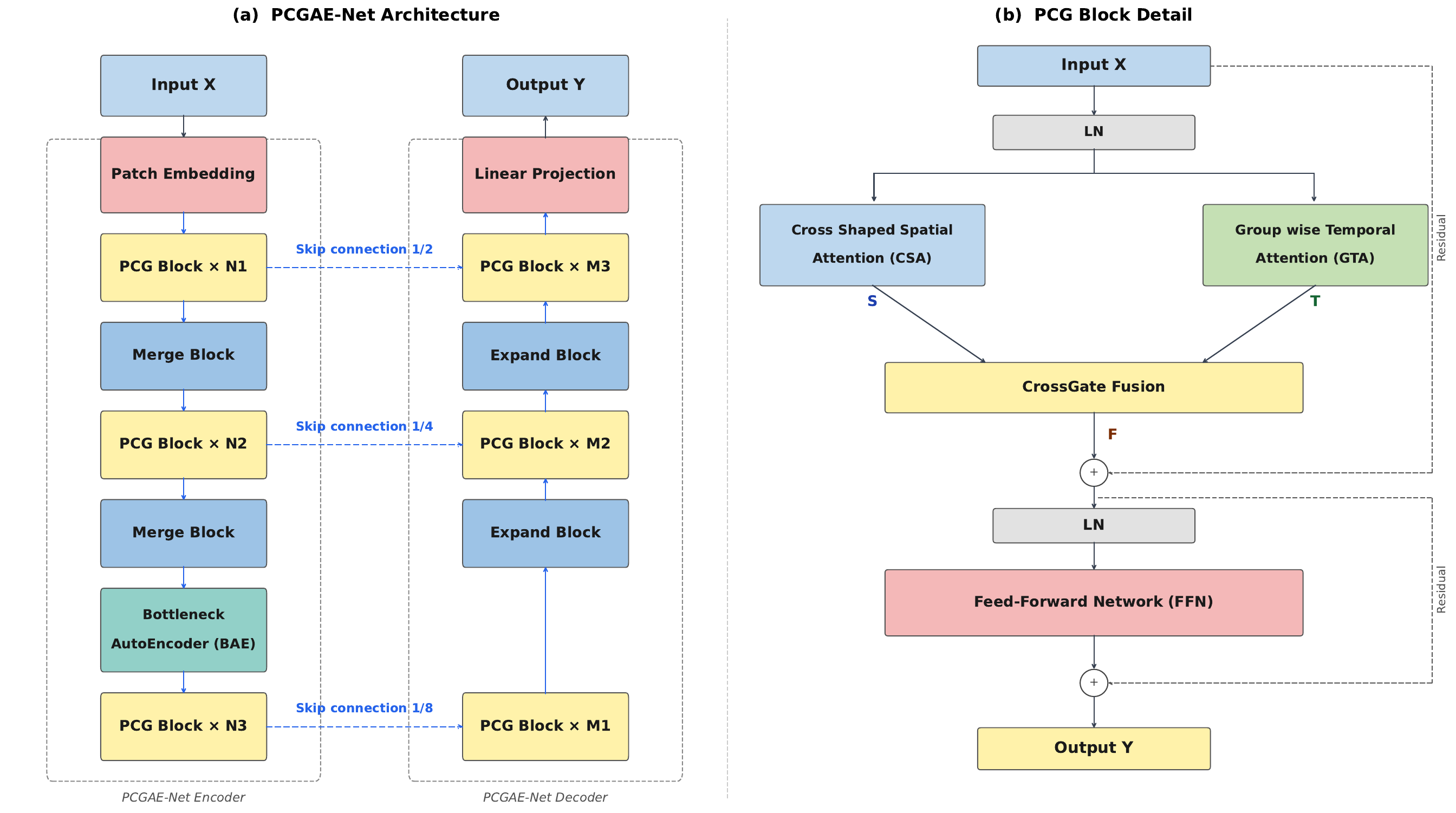}
\caption{Proposed architecture. (a) Lightweight PCGAE-Net: a two-level U-Net encoder-decoder with PCG Attention Blocks at each skip level and a BAE at the deepest stage. Skip connections carry encoder features to the symmetric decoder levels. (b) PCG Block: CSA and GTA receive the same layer-normalized input independently. Their outputs $\mathbf{S}$ and $\mathbf{T}$ are fused by a per-channel sigmoid CrossGate, added to the residual stream, and processed by an FFN.}
\label{fig:architecture}
\vspace{-2ex}
\end{figure*}

The overall structure of Lightweight PCGAE-Net follows a U-Net encoder-decoder, illustrated in Fig.~\ref{fig:architecture}. The encoder progressively downsamples angle and delay dimensions using convolutional merge blocks, building a feature hierarchy. The decoder symmetrically upsamples with expand blocks and skip connections from corresponding encoder levels. Three modifications distinguish our design from CS3T-UNet: the Parallel CrossGate (PCG) Attention Block replaces sequential attention at each skip level, the Bottleneck AutoEncoder (BAE) replaces vanilla attention at the deepest layer, and overall depth is reduced from three merge/expand pairs to two.

\subsection{Parallel CrossGate Attention Block}
\label{subsec:crossgate}

CS3T-UNet applies CSA and GTA sequentially at each encoder-decoder level:
\begin{equation}
\mathbf{X}' = \mathbf{X} + \mathrm{CSA}(\mathrm{LN}(\mathbf{X})), \quad
\mathbf{X}'' = \mathbf{X}' + \mathrm{GTA}(\mathrm{LN}(\mathbf{X}')).
\label{eq:seq}
\end{equation}
CSA computes cross-shaped attention across spatial positions and adds it to the residual stream. GTA then applies group-wise temporal attention to the already-modified features $\mathbf{X}'$ rather than to the original $\mathbf{X}$. In slow-fading conditions, this ordering has little effect; as UE speed increases and temporal channel variation dominates, the temporal signatures in the original features that GTA needs to capture become harder to recover from the spatially-processed $\mathbf{X}'$.

The PCG Attention Block fixes this by routing both modules to the same input:
\begin{equation}
\mathbf{S} = \mathrm{CSA}(\mathrm{LN}(\mathbf{X})), \quad
\mathbf{T} = \mathrm{GTA}(\mathrm{LN}(\mathbf{X})).
\label{eq:parallel}
\end{equation}
CSA and GTA now operate independently on the original channel representation, producing feature maps $\mathbf{S}$ and $\mathbf{T}$ respectively. These are combined through a CrossGate---a learned per-channel sigmoid weighting:
\begin{equation}
g = \sigma\!\left(\mathbf{W}_g\,[\mathbf{S};\mathbf{T}]+\mathbf{b}_g\right), \quad
\mathbf{F} = g\odot\mathbf{S} + (1-g)\odot\mathbf{T},
\label{eq:gate}
\end{equation}
where $[\mathbf{S};\mathbf{T}]\!\in\!\mathbb{R}^{2C}$ is the channel-dimension concatenation, $\mathbf{W}_g\!\in\!\mathbb{R}^{C\times 2C}$, $\sigma(\cdot)$ is the sigmoid function, and $\odot\mathbf{S}$ denotes element-wise multiplication. The gate $g\!\in\!(0,1)^C$ produces a separate weight for each channel dimension, letting different feature channels favor spatial or temporal information as the channel conditions require. We initialize $\mathbf{W}_g\!=\!\mathbf{0}$ and $\mathbf{b}_g\!=\!\mathbf{0}$, giving $g\!=\!0.5$ at training start---an equal mixture of the two streams. Starting from this balanced point avoids early-training instability that can arise when one stream dominates before the optimizer has found a good feature space. In practice, the gate values learned after training differ substantially from 0.5 in a channel-specific way, confirming that the gate does useful work.

The full block output includes a feed-forward network (FFN) and a residual connection:
\begin{equation}
\mathbf{Y} = \mathbf{X} + \mathbf{F} + \mathrm{FFN}(\mathrm{LN}(\mathbf{X}+\mathbf{F})).
\label{eq:ffn}
\end{equation}
The CrossGate itself adds only $2C^2\!+\!C$ parameters per block---negligible relative to the attention and FFN modules. The architectural benefit of independent feature extraction comes at essentially no parameter cost.

\subsection{Bottleneck AutoEncoder}
\label{subsec:bae}

In the deepest encoder stage, feature depth has grown to $8C$ through the successive merge operations. CS3T-UNet applies full cross-shaped attention at this resolution, which is the single most expensive operation in the network. Many of these $8C$ feature channels at the bottleneck are redundant: they capture global statistics that summarize information already encoded in the encoder's shallower representations.

The Bottleneck AutoEncoder replaces this with a compression-decompression block:
\begin{align}
\mathbf{Z} &= \mathrm{GELU}\!\left(\mathrm{GN}\!\left(\mathrm{Conv}_{1\times1}(\mathbf{X})\right)\right), \label{eq:bae_enc} \\
\hat{\mathbf{X}} &= \mathrm{GN}\!\left(\mathrm{Conv}_{1\times1}(\mathbf{Z})\right), \quad \mathbf{Y}_{\mathrm{BAE}} = \hat{\mathbf{X}} + \mathbf{X}. \label{eq:bae_dec}
\end{align}
The encoder convolution maps from $8C$ to $4C$ channels; the decoder restores the original depth. Group Normalization (GN) is applied after each convolution, GELU provides the nonlinearity, and a residual connection ensures gradient flow. The compressed features $\mathbf{Z}\!\in\!\mathbb{R}^{4C}$ are passed to a smaller attention module operating at half the original bottleneck depth, cutting the attention cost at this stage by approximately 75\%.

Training minimizes a combined loss:
\begin{equation}
\mathcal{L} = \mathcal{L}_{\mathrm{pred}} + \alpha\,\mathcal{L}_{\mathrm{rec}},
\label{eq:loss}
\end{equation}
where $\mathcal{L}_{\mathrm{pred}}$ is the NMSE prediction loss and $\mathcal{L}_{\mathrm{rec}}\!=\!\|\hat{\mathbf{X}}-\mathbf{X}\|_F^2$ is the bottleneck reconstruction loss. We use $\alpha\!=\!0.1$. The reconstruction loss acts as a self-supervised regularizer: it penalizes compression that discards features the decoder will need, even when those features do not directly reduce the current prediction loss on the training set. This is particularly valuable when combining the BAE with aggressive channel-dimension reduction (from $C\!=\!64$ to $C\!=\!48$), where the compressed representation must work harder to preserve the spatial-temporal structure needed for accurate multi-step prediction.

\subsection{Slim Encoder-Decoder and Dimensionality Reduction}
\label{subsec:slim}

Beyond the attention-level changes, we make two structural modifications to reduce parameter count. First, the number of merge/expand block pairs is reduced from three to two. The third pair in CS3T-UNet contributes relatively little to prediction accuracy while adding roughly 16\,M parameters,and the remaining accuracy gap is largely recovered through improved attention quality from the CrossGate. Second, we apply delay-domain cropping to the $N_f\!=\!32$ most energetic ADP bins, reducing input size by half along the delay domain. The discarded 32 bins carry mostly noise at the carrier and UE speeds we consider. Combined with reducing base channel width from $C\!=\!64$ to $C\!=\!48$,  these changes bring the total parameter count from 31.74\,M (full PCGAE-Net) to 8.54\,M.

Table~\ref{tab:config} summarizes all model configurations evaluated in this work.

\begin{table}[t]
\centering
\caption{Model configurations. Depth = No. of merge/expand block pairs.}
\label{tab:config}
\resizebox{\columnwidth}{!}{%
\begin{tabular}{lcccc}
\toprule
Model & Depth & $C$ & $N_f$ & BAE \\
\midrule
CS3T-UNet (baseline)        & 3 & 64 & 64 & No  \\
PCGAE (w/o AE)              & 3 & 64 & 64 & No  \\
PCGAE (Full)                & 3 & 64 & 64 & Yes \\
PCGAE (DR, $C\!=\!48$)      & 3 & 48 & 32 & Yes \\
PCGAE (Slim)                & 2 & 64 & 64 & Yes \\
\textbf{Proposed (Slim+DR)} & \textbf{2} & \textbf{48} & \textbf{32} & \textbf{Yes} \\
\bottomrule
\end{tabular}}
\end{table}

\section{Experimental Setup}
\label{sec:exp}

\subsection{Dataset and Simulation Conditions}

All experiments use channels generated by the QuaDriGa simulator~\cite{jaeckel2014quadriga} under the 3GPP 38.901 UMa NLOS propagation model at 5\,GHz carrier frequency, with $N_t\!=\!64$ BS antennas and $N_f\!=\!64$ OFDM subcarriers. After transformation to the ADP domain, $N_f$ and $N_t$ correspond to delay bins and angle bins respectively, before any dimensionality reduction. Three UE speeds are evaluated: 5, 9, and 12\,km/h, representing pedestrian, slow vehicular, and moderate vehicular mobility.Each dataset contains 9,000 training sequences and 1,000 test sequences. Each sample is a sequence of 20 ADP-domain channel snapshots, over which a sliding window of size T + L is applied, yielding more independent windows for L = 1 than for L = 5. The first T = 10 frames of each window serve as input history and the remaining L frames as prediction targets. The input tensor shape before delay domain cropping is $(20,2,64,64)$. 

The three speeds probe a meaningful range of Doppler conditions, while keeping experiments tractable. At 5\,km/h the channel evolves slowly and spatial modeling dominates; at 12\,km/h temporal variation becomes the dominant factor and 5-step prediction is substantially harder for all methods. The speed of 9\,km/h sits at an intermediate point where spatial and temporal dynamics matter, and it turns out to be where the parallel CrossGate provides its largest benefit.

\subsection{Training Protocol}

All models are trained with AdamW at an initial learning rate of $2\!\times\!10^{-4}$ with decaying cosine annealing to $10^{-6}$ over 400 epochs with batch size 32. BAE models are trained using the combined loss in~(\ref{eq:loss}) with $\alpha\!=\!0.1$; others use $\mathcal{L}_{\mathrm{pred}}$ alone. All experiments run on an NVIDIA RTX 6000 Ada GPU in bfloat16 precision without data augmentation.

The weight initialization for CrossGate parameters follows the convention in Section~\ref{subsec:crossgate}: $\mathbf{W}_g\!=\!\mathbf{0}$ and $\mathbf{b}_g\!=\!\mathbf{0}$. All other parameters use the default PyTorch scheme.

All models share the same training protocol; parameter counts and configuration details are given in Table~\ref{tab:config}.

\section{Results and Analysis}
\label{sec:results}

\subsection{Main Comparison}

\begin{figure}[t]
\centering
\includegraphics[width=\columnwidth]{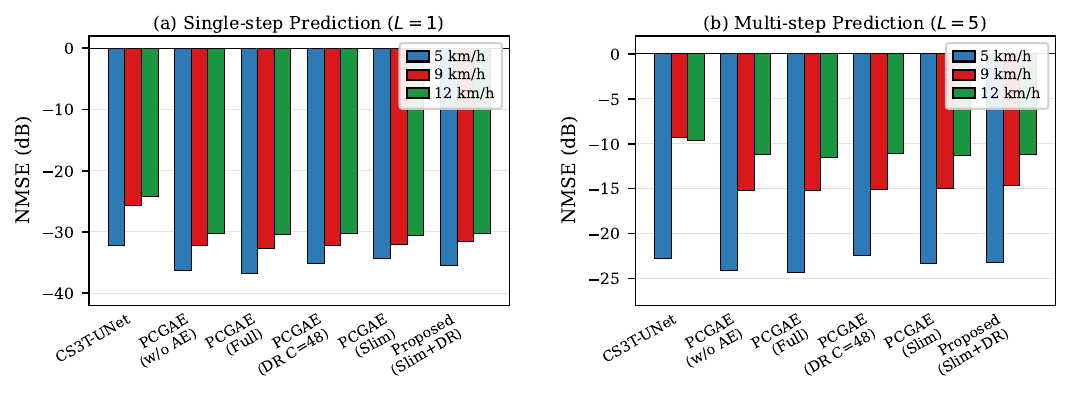}
\caption{NMSE (dB) for all models at 5, 9, and 12\,km/h. Bars are grouped by model; three bars per group correspond to three UE speeds. Lower is better. The proposed model consistently outperforms CS3T-UNet across all conditions despite using 58\% fewer parameters.}
\label{fig:nmse_compare}
\end{figure}

\begin{figure}[t]
\centering
\includegraphics[width=\columnwidth]{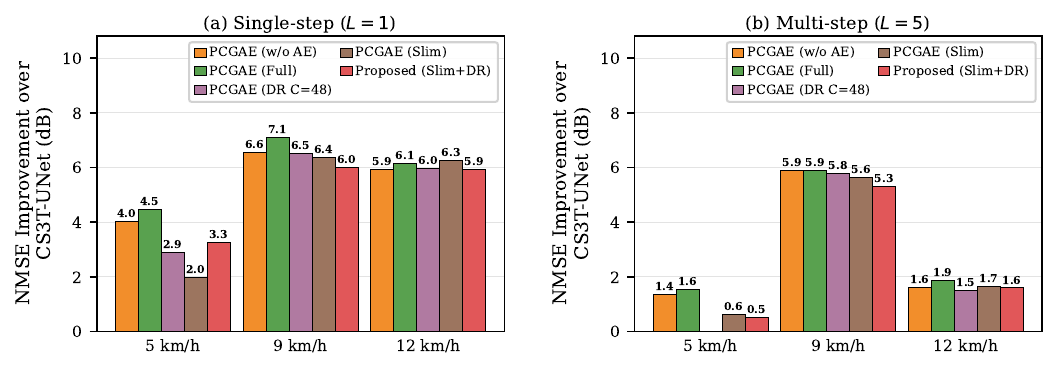}
\caption{NMSE improvement (dB) over CS3T-UNet baseline, grouped by UE speed. At $L\!=\!1$, the proposed model achieves 3.3--6.0\,dB gain depending on speed; at $L\!=\!5$, gains range from 0.5 to 5.9\,dB.}
\label{fig:improvement}
\end{figure}

\begin{figure}[t]
\centering
\includegraphics[width=\columnwidth]{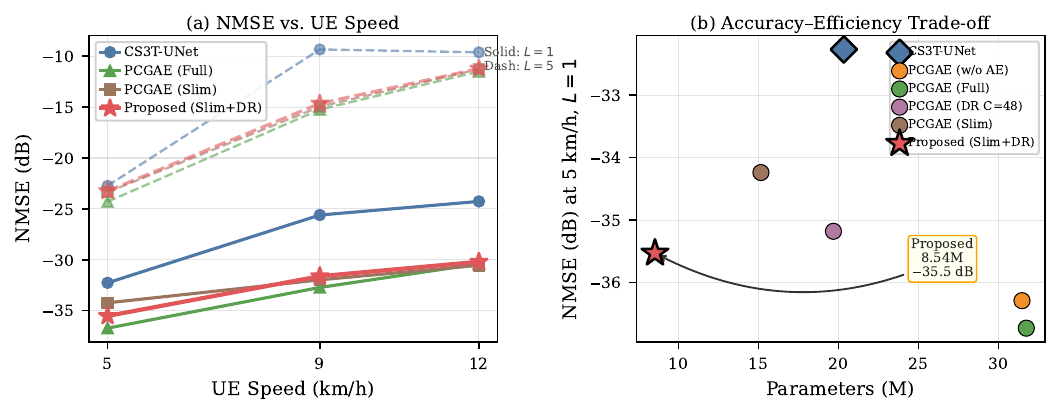}
\caption{(a) NMSE vs.\ UE speed for key model variants. Solid: $L\!=\!1$; dashed: $L\!=\!5$. (b) Parameter count vs.\ NMSE at 5\,km/h, $L\!=\!1$. The proposed model (starred) achieves the best accuracy-per-parameter.}
\label{fig:tradeoff}
\end{figure}

Table~\ref{tab:main} reports NMSE for all models at all speed-horizon combinations. Fig.~\ref{fig:nmse_compare} visualizes the same data as a grouped bar chart, Fig.~\ref{fig:improvement} shows per-speed NMSE improvement over CS3T-UNet, and Fig.~\ref{fig:tradeoff} summarizes the accuracy-efficiency trade-off.

\begin{table}[t]
\centering
\caption{NMSE(dB). Speed columns in km/h. Lower is better. \textbf{Bold}: best.}
\label{tab:main}
\resizebox{\columnwidth}{!}{%
\setlength{\tabcolsep}{4pt}
\small
\begin{tabular}{lrcccccc}
\toprule
& & \multicolumn{3}{c}{$L=1$} & \multicolumn{3}{c}{$L=5$} \\
\cmidrule(lr){3-5}\cmidrule(lr){6-8}
Model & Params & 5 & 9 & 12 & 5 & 9 & 12 \\
\midrule
GRU              & 4.99\,M  & $-3.6$  & $-3.2$  & $-3.1$  & $-0.3$  & $+0.0$  & $+0.0$  \\
CS3T-UNet        & 20.34\,M & $-32.3$ & $-25.6$ & $-24.3$ & $-22.7$ & $-9.3$  & $-9.6$  \\
PCGAE (w/o AE)   & 31.47\,M & $-36.3$ & $-32.2$ & $-30.2$ & $-24.1$ & $-15.2$ & $-11.2$ \\
PCGAE Full       & 31.74\,M & $\mathbf{-36.7}$ & $\mathbf{-32.7}$ & $\mathbf{-30.4}$ & $\mathbf{-24.3}$ & $\mathbf{-15.2}$ & $-11.5$ \\
PCGAE DR ($C{=}48$) & 19.69\,M & $-35.2$ & $-32.1$ & $-30.3$ & $-22.4$ & $-15.1$ & $-11.1$ \\
PCGAE Slim       & 15.16\,M & $-34.2$ & $-32.0$ & $-30.6$ & $-23.4$ & $-15.0$ & $\mathbf{-11.3}$ \\
\textbf{Proposed (Slim+DR)} & \textbf{8.54\,M} & $-35.5$ & $-31.6$ & $-30.2$ & $-23.2$ & $-14.6$ & $-11.2$ \\
\bottomrule
\end{tabular}}
\end{table}

The GRU baseline confirms that the prediction task is genuinely non-trivial: it reaches only $-3.57$\,dB at 5\,km/h ($L\!=\!1$), nearly 29\,dB below the transformer-based methods. This gap is not a parameter-count issue---GRU has 4.99\,M parameters, not far below the proposed model at 8.54\,M---but reflects the fundamental limitation of recurrent networks in jointly capturing spatial-temporal structure of MIMO channels.

Among transformer-based models, the proposed method is an interesting position. It outperforms the 20.34\,M CS3T-UNet baseline at every speed and horizon despite using 58\% fewer parameters, and it comes within 1.2\,dB of the 31.74\,M PCGAE (Full) while being less than a quarter its size. No competing configuration achieves this combination.

Furthermore, the inference time is almost the same across all model variants, confirming that the architectural changes do not introduce additional latency overhead despite the difference in parameter counts.

\subsection{Effect of UE Speed}

In single-step prediction ($L\!=\!1$), the gain from parallel CrossGate grows with speed. The improvement from PCGAE (w/o AE) over CS3T-UNet increases from 4.0\,dB at 5\,km/h to 6.6\,dB at 9\,km/h. The simplest explanation is that as temporal variation becomes the dominant channel characteristic, freeing GTA from operating on spatially-preprocessed features matters more. At 5\,km/h the channel changes slowly enough that the bias in CS3T-UNet's sequential design has limited impact; at 9\,km/h it costs substantially more. The proposed model tracks this trend: 3.3\,dB gain at 5\,km/h and 6.0\,dB at 9\,km/h.

In multi-step prediction ($L\!=\!5$), gains at 9\,km/h are 5.9\,dB for PCGAE (Full) and 5.3\,dB for the proposed model over the baseline. The baseline itself drops sharply from $-22.73$\,dB at 5\,km/h to $-9.33$\,dB at 9\,km/h as the 5-step prediction becomes harder under the Doppler effect; the PCGAE variants maintain prediction quality much better in this regime, which is where the practical benefit of parallel architecture is most evident.

\subsection{Ablation Study}

Table~\ref{tab:ablation} isolates each design choice by adding one modification at a time, evaluated at 5\,km/h ($L\!=\!1$) to avoid confounding with speed-dependent effects.

\begin{table}[t]
\centering
\caption{Ablation study at 5\,km/h, $L\!=\!1$. Each row adds one design change to the previous.}
\label{tab:ablation}
\resizebox{\columnwidth}{!}{%
\begin{tabular}{lrcc}
\toprule
Configuration & Params & NMSE (dB) & $\Delta$ vs.\ CS3T-UNet \\
\midrule
CS3T-UNet (baseline)                & 20.34\,M & $-32.27$ & --- \\
Parallel CrossGate (w/o BAE)      & 31.47\,M & $-36.29$ & $+4.02$\,dB \\
+ Bottleneck AutoEncoder            & 31.74\,M & $-36.73$ & $+4.46$\,dB \\
+ Slim encoder-decoder (2 levels)   & 15.16\,M & $-34.24$ & $+1.97$\,dB \\
+ DR ($C\!=\!48$, $N_f\!=\!32$)     & \textbf{8.54\,M}  & $-35.53$ & $+3.26$\,dB \\
\bottomrule
\vspace{-2ex}
\end{tabular}}
\end{table}

Parallel CrossGate alone provides the largest single improvement (4.02\,dB), confirming that the sequential ordering dependency is the primary structural limitation. The BAE then contributes a further 0.44\,dB at a cost of only 0.27\,M additional parameters. This small gain has an outsized effect on what compression is viable: as will be apparent from the next row, reducing depth without the reconstruction regularizer causes a larger accuracy drop than reported because the bottleneck discards channel structure that the decoder needs to recover at the skip connections.

Reducing depth from three levels to two costs 2.49\,dB relative to the full model---the largest single accuracy loss. At 15.16\,M parameters, however, even this reduced model is already 1.97\,dB better than the baseline. Adding dimensionality reduction then recovers 1.29\,dB at a saving of 6.62\,M parameters, producing the final proposed configuration.

\subsection{Discussion}

Two broader observations are worth noting. The first is that architectural correctness matters more than scale. PCGAE (Full) has 56\% more parameters than CS3T-UNet, yet the gain from fixing the sequential attention ordering (4.02\,dB from the parallel CrossGate alone) is larger than what any parameter increase alone could plausibly explain. The structural bias is a genuine limitation, not a capacity issue, and correcting it first---before pursuing scale reduction---is what allows the proposed model to stay well above the baseline despite being less than half its size.

The second observation concerns the role of the auxiliary reconstruction loss in the BAE. As shown in Table~\ref{tab:ablation}, applying dimensionality reduction without the BAE leads to a noticeable accuracy drop, since prediction loss alone provides no explicit signal to preserve channel structure at the bottleneck. The reconstruction loss addresses this by directly regularizing the compressed representation, giving the model a second training objective that specifically rewards the retention of useful information for the decoder. This becomes especially important under aggressive compression, where the channel width is reduced from C = 64 to C = 48, halving the frequency dimension. The benefit is clearly reflected in Table~\ref{tab:ablation}: the final row recovers 1.29 dB at a saving of 6.62 M parameters, which is largely attributable to the reconstruction loss keeping the bottleneck representation informative under such compression.
\section{Conclusion}
\label{sec:conclusion}

This paper presented Lightweight PCGAE-Net, a 5G channel predictor built around two targeted architectural corrections to CS3T-UNet. Running cross-shaped spatial and group-wise temporal attention in parallel on the same input, with a learned per-channel CrossGate to combine their outputs, removes the sequential ordering bias that constrains GTA's ability to capture temporal channel dynamics---an effect that grows more significant as UE speed increases. A Bottleneck AutoEncoder at the deepest encoder stage compresses feature depth by half, cutting the most expensive self-attention operation, while an auxiliary reconstruction loss prevents the compressed representation from losing prediction-relevant structure. Combining these with a shallower encoder-decoder and frequency-domain input cropping yields a model with 8.54\,M parameters that outperforms the 20.34\,M CS3T-UNet baseline by 3.3--6.0\,dB across all evaluated conditions on QuaDriGa.

Future directions include extending evaluation to higher mobility regimes (30-120 km/h) would validate the CrossGate's robustness under severe Doppler conditions representative of vehicular and railway communication scenarios.

\bibliographystyle{IEEEtran}
\bibliography{references}

\end{document}